\documentclass[aps,prl,reprint,amsmath,amssymb,showpacs,superscriptaddress,twocolumn]{revtex4}
\usepackage{graphicx}
\usepackage{xcolor}

\usepackage{subfigure}
\usepackage{hyperref,hypcap}
\usepackage{braket}
\usepackage{amsmath}
\usepackage{commath}

\begin{document}
	\title{Time-Reversal Symmetry-Breaking Nematic Insulators \\
		near Quantum Spin Hall Phase Transitions}
	\author{Fei Xue}\email{feixue@utexas.edu}
	\author{A.H. MacDonald}\email{macd@physics.utexas.edu}
	\address{Department of Physics, University of Texas at Austin, Austin TX 78712}
	
	\date{\today}
	
	\begin{abstract}
		We study the phase diagram of a model quantum spin Hall system as a function of band 
		inversion and band-coupling strength, demonstrating that when band hybridization is weak,
		an interaction-induced nematic insulator state emerges over a wide range of band 
		inversion. This property is a consequence of the long-range Coulomb interaction,
		which favors interband phase coherence that is weakly dependent on 
		momentum and therefore frustrated by the single-particle Hamiltonian at the band inversion point.
		For weak band hybridization, interactions convert the continuous gap closing topological phase transition at 
		inversion into a pair of continuous phase transitions bounding a state with 
		broken time-reversal and rotational symmetries.  At intermediate band hybridization,
		the topological phase transition proceeds instead via a quantum anomalous Hall insulator state, whereas 
		at strong hybridization interactions play no role.  
		We comment on the implications of our findings for InAs/GaSb and HgTe/CdTe quantum spin Hall systems.  	
	\end{abstract}
	
	\maketitle

	
	\begin{figure}[b]
		\includegraphics[width=1\columnwidth]{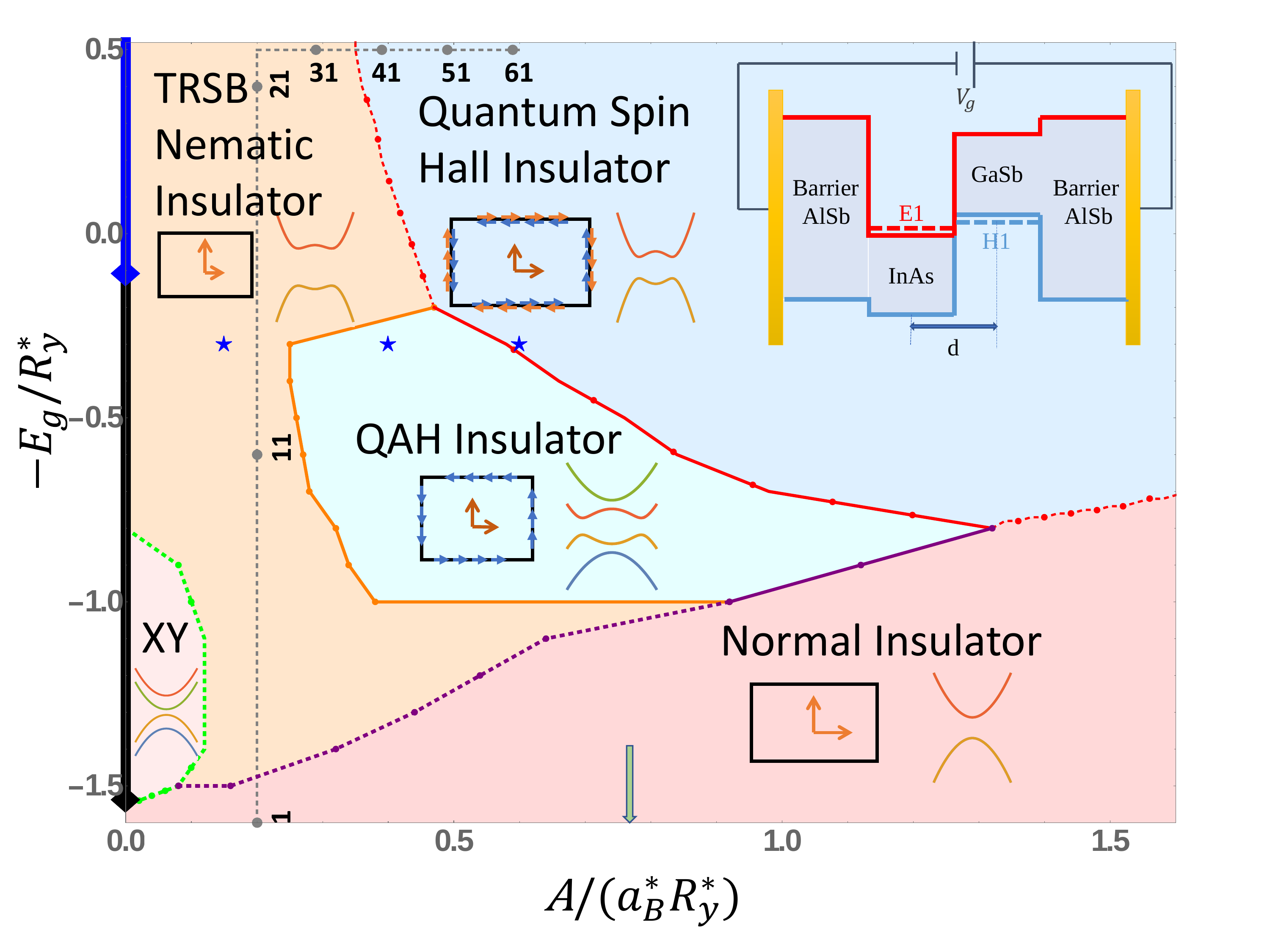}
		\caption{(Color online) 
			Mean-field phase diagram of a model quantum spin Hall insulator (QSHI) as a function of the band inversion
			parameter $E_{g}$ and the band hybridization parameter $A$.  
			The black diamond on the $A=0$ axis separates a normal insulator and an excitonic insulator (bold black),
			and the blue diamond marks the first-order Mott transition between an excitonic insulator and 
			a metallic state (bold blue) that occurs at large exciton density. 
			The exciton condensate state is characterized by spontaneous phase coherence 
			between conduction and valence bands, and therefore exists only along the $A=0$ line.   
			The stability regions of the five finite $A$ states we have identified are 
			distinguished by color [blue for the QSHI, cyan for the quantum anomalous Hall insulator(QAHI), 
			orange for the time-reversal symmetry-breaking(TRSB) nematic insulator, pink for the 
			TRSB nematic insulator state with an additional $XY$ broken symmetry, 
			and red for the normal insulator state].  Each state is distinguished by its typical dressed band 
			structure, and by the presence or absence of edge states which is indicated using schematic Hall bars.
			Solid lines indicate first-order phase transitions and dashed lines indicate continuous phase transitions.
			The gray dashed line is the path connecting the normal insulator and QSHI state
			via the TRSB nematic insulator state discussed in the main text and illustrated in 
			Fig.~\ref{Fig:circle}, and the gray circles correspond to every tenth point plotted in that figure.
			The blue stars specify the phase diagram points at which we illustrate quasiparticle band topological properties in Fig.~\ref{Fig:Wilsonloop}.
			The inset shows a schematic band diagram for the AlSb/InAs/GaSb/AlSb QW system, to which the model corresponds most closely.  
			The arrow along the horizontal axis indicates the value of the dimensionless band hybridization 
			parameter for the case of adjacent InAs and GaSb layers.
		}
		\label{Fig:PhaseDiagram}
	\end{figure}

	{\em Introduction.}---
	The quantum spin Hall insulator (QSHI) is a state of two-dimensional matter that supports
	gapless helical edge modes protected by time-reversal symmetry\cite{Kane2005,Kane2005_2,Bernevig2006,Liu2008}.
	Recent experiments in high-quality HgTe/CdTe\cite{Konig2007,Kononov2015,Molenkamp2016,Molenkamp2017,Ganichev2017} and type-II InAs/GaSb\cite{Du2011,Kouwenhoven2015,Du2015,Du2017Strained,Kononov2017}
	quantum wells (QWs) have demonstrated that phase transitions between normal insulators and
	QSHIs can be generated in QW systems by engineering a band crossing between 
	conduction and heavy-hole bands.  In this Letter we show that
	when band hybridization at finite momenta is weak, electron-electron
	interactions can alter the character of these transition by inserting an intermediate gapped electron 
	nematic insulator state between the normal insulator and QSHI states(NI/QSHI).
	Our principle results are summarized in Fig.~\ref{Fig:PhaseDiagram} in which we distinguish 
	five phases, including a normal insulator with a full valence band and an empty conduction band,
	a QSHI with inverted bands at $\vec{k}=0$ and an avoided crossing 
	gap at finite momentum, a nematic insulator state in which both rotational symmetry and time-reversal
	symmetry are broken, an $XY$ insulator, and a quantum anomalous Hall insulator(QAHI).
	We explain why the nematic state is made inevitable by the large energy difference 
	between $s$-wave and $p$-wave Wannier excitons in two dimensions, and by the tendency of dilute excitons
	with repulsive interactions to condense\cite{Keldysh1965,Lozovik1976,Comte1982}.

	Although our conclusions are quite general, 
	the detailed calculations described below employ a four-band Bernevig-Hughes-Zhang (BHZ) 
	model\cite{Bernevig2006,Liu2008}, which describes inversion between quantum well $s$ and 
	heavy-hole states.  We neglect bulk inversion asymmetry and structural inversion asymmetry terms in
	the band Hamiltonian because they are normally small\cite{Liu2013}.  
	With this approximation the BHZ model separates into time-reversed diagonal blocks. 
	In the basis 
	$\{\ket{E_1\uparrow},\ket{H_1\uparrow},\ket{E_1\downarrow},\ket{H_1\downarrow}\}$, the QW band Hamiltonian is  
	\begin{equation}
	\label{eq:Hamiltonian}
	\hat{H}=\sum_{\vec{k}}\psi_{\vec{k}}^{\dagger}
	\begin{pmatrix}
	H_{0,\uparrow} & 0\\
	0 & H_{0,\downarrow}\\
	\end{pmatrix}
	\psi_{\vec{k}}
	+\hat{H}_I,
	\end{equation}
	where the down spin single-particle term is 
	\begin{equation}
	\label{eq:single_Hamiltonian}
	H_{0,\downarrow}=\begin{pmatrix}
	\frac{\hbar^2 k^2}{2m_e}+E_c & -Ak_- \\
	-Ak_+ & -\frac{\hbar^2 k^2}{2m_h}+E_v \\
	\end{pmatrix},
	\end{equation}
	where $k_{\pm}=k_x\pm \mathrm{i} k_y$,  $A$ is the band hybridization strength, 
	and $m_{e(h)}$ is the electron(hole) effective mass.
	Our study is motivated by recent experimental\cite{Du2015,Yu2017} 
	and theoretical\cite{Budich2014,Hyart2014} work 
	that has demonstrated that interactions can play an essential role near NI/QSHI phase 
	transitions.  The Coulombic electron-electron interaction Hamiltonian is
	\begin{equation}
	\label{eq:interaction_Hamiltonian}
	\hat{H}_I=\frac{1}{2S}\sum_{\sigma\sigma',ss'}\sum_{\vec{k},\vec{k}',\vec{q}} \, V^{ss'}(\vec{q}) \, 
	a_{\sigma s \vec{k}}^{\dagger}a_{\sigma' s' \vec{k}'}^{\dagger}
	a_{\sigma' s' \vec{k}'+\vec{q}}a_{\sigma s \vec{k}-\vec{q}} \, 
	\end{equation}
	where $S$ is the two-dimensional system area, 
	$s(s')=c$ (conduction) or $v$ (valence) and $\sigma(\sigma')=\uparrow$ or $\downarrow$ 
	distinguish band and spin states, $a_{\sigma s \vec{k}}^{\dagger}$ and 
	$a_{\sigma s \vec{k}}$ are creation and annihilation operators, 
	$V^{cc}(\vec{q})=V^{vv}(\vec{q})=V(\vec{q})=2\pi e^2/(\epsilon q)$,
	$V^{cv}(\vec{q})=V^{vc}(\vec{q})=U(\vec{q})=V(\vec{q}) e^{-qd}$, and $d$ is the 
	spatial separation between conduction and valence band layers.
	We are interested in the properties of this interacting electron 
	system as the band gap $E_{g}=E_c-E_v$ 
	closes and changes sign.

	When $A$ vanishes, the model reduces to that of the well-understood
	two-dimensional excitonic insulator problem\cite{Zhu1995,Lozovik1996,Fernandez1996,Vina1999,
		Combescot2008a,Combescot2008b,Butov2012,Perali2013,Wu&Xue2015},
	which features a continuous phase transition between a trivial band insulator 
	and a state that is still insulating but populated by a condensate of excitons with weakly 
	repulsive interactions.  The phase transition occurs not at $E_g=0$, but at a positive 
	$E_g$ value equal to the exciton binding energy.
	The characteristic length scale of
	the excitonic insulator problem is the effective Bohr radius
	$a_B^*=\epsilon \hbar^2  /(me^2)$, and the 
	characteristic energy scale is the effective 
	Rydberg $\text{Ry}^*=e^2/(2\epsilon a_B^*)$.
	[Here, $m=m_em_h/(m_e+m_h)$ is the excitonic reduced mass.]
	We explain below how excitonic insulator physics 
	evolves with increasing $A/(Ry^* a_B^*)$ into a renormalized version of 
	a single-particle NI/QSHI phase-transition physics, and why the crossover as $A$ is varied
	involves a state with broken rotational and time-reversal symmetry. Some aspects of the 
	physics are best illustrated using a simplified two-band model, whose properties
	are discussed in detail in the Supplemental Material.  
	
	{\em Microscopic mean-field theory.}---
	We first describe the results of a mean-field theory calculation that allows for 
	all possible broken symmetries that preserve translational invariance, and then 
	discuss how neglected quantum fluctuations might alter the resulting phase diagram.  
	The Hartree-Fock mean-field Hamiltonian for the BHZ model is
	
	\begin{equation}
	\label{eq:MF}
	\hat{H}_{MF}=\sum_{\vec{k}} \psi_{\vec{k}}^{\dagger}
	(H_0+H_{Hartree}+H_{Fock})
	\psi_{\vec{k}}
	\end{equation}
	where 
	\begin{equation}
	\label{eq:exchange}
	H_{Fock}=
	\begin{pmatrix}
	\Delta_{\uparrow\uparrow}^{cc}(\vec{k}) &\Delta_{\uparrow\uparrow}^{cv}(\vec{k})& \Delta_{\uparrow\downarrow}^{cc}(\vec{k}) & \Delta_{\uparrow\downarrow}^{cv}(\vec{k})\\
	\Delta_{\uparrow\uparrow}^{vc}(\vec{k})&\Delta_{\uparrow\uparrow}^{vv}(\vec{k}) &  \Delta_{\uparrow\downarrow}^{vc}(\vec{k}) & \Delta_{\uparrow\downarrow}^{vv}(\vec{k})\\
	\Delta_{\downarrow\uparrow}^{cc}(\vec{k}) &\Delta_{\downarrow\uparrow}^{cv}(\vec{k})& \Delta_{\downarrow\downarrow}^{cc}(\vec{k}) &  \Delta_{\downarrow\downarrow}^{cv}(\vec{k})\\
	\Delta_{\downarrow\uparrow}^{vc}(\vec{k})& \Delta_{\downarrow\uparrow}^{vv}(\vec{k})&  \Delta_{\downarrow\downarrow}^{vc}(\vec{k}) & \Delta_{\downarrow\downarrow}^{vv}(\vec{k}),
	\end{pmatrix},
	\end{equation}
	\begin{equation}
	\label{eq:band}
	H_0+H_{Hartree}=\zeta_{\vec{k}}s_0 \tau_0+\epsilon_{\vec{k}}s_0 \tau_z+A k_xs_z\tau_x-A k_ys_0\tau_y,
	\end{equation} 
	and $s_{i}$ and $\tau_{i}$ are spin and electron-hole Pauli matrices respectively.
	In Eq.~\ref{eq:band} $\zeta_{\vec{k}}=\hbar^2k^2[1/(4m_e)-1/(4m_h)]$ accounts for the mass difference 
	between conduction and valence bands, which plays a minor
	role and is dropped below.  
	The band-splitting term, $\epsilon_{\vec{k}}=\hbar^2k^2/4m+E_g/2+2\pi e^2 n_{ex}d$, includes an 
	electrostatic Hartree contribution which is linear in $d$.  In Eq.~\ref{eq:exchange}
	\begin{equation}
	\label{eq:SC}
	\Delta_{\sigma\sigma'}^{ss'}(\vec{k})=-\frac{1}{S}\sum_{\vec{k}'}V^{ss'}(\vec{k}-\vec{k}')
	\rho_{\sigma\sigma'}^{ss'}(\vec{k}'),
	\end{equation}
	where the density matrix,
	\begin{equation}
	\label{eq:rho}
	\rho_{\sigma\sigma'}^{ss'}(\vec{k})=\braket{a_{\sigma's' \vec{k}}^{\dagger} a_{\sigma s \vec{k}}}
	-\delta_{ss'}\delta_{\sigma \sigma'}\delta_{\sigma=v},
	\end{equation}  
	is defined relative to the fully filled valence band because the bare bands are assumed to be 
	those of the normal insulator.
	The exciton density appearing in the Hartree term is 
	\begin{equation}
	n_{ex}=\frac{1}{S}\sum_{\sigma,\vec{k}}\rho_{\sigma\sigma}^{cc}(\vec{k}) .
	\end{equation}
	In much of the phase diagram Eq.~\ref{eq:MF} has multiple metastable solutions.  We 
	select the mean-field ground state by computing the total energy per area:
	\begin{equation}
	\label{eq:totalenergy}
	\epsilon=\frac{1}{2S}\sum_{\vec{k}}Tr\{\rho(\vec{k}) \, [H_0(\vec{k})+H_{MF}(\vec{k})] \}.
	\end{equation}
	
	The BHZ single-particle Hamiltonian $H_0$ is isotropic and has time-reversal symmetry. 
	Its coupling between the $s$-wave conduction and $p$-wave valence bands vanishes 
	at $\vec{k}=0$ because of an underlying microscopic $C4$ rotational symmetry.
	It follows that 
	rotational symmetry is broken when $\Delta_{\sigma\sigma'}^{cv}(\vec{k}=0)\ne 0$, 
	allowing us to identify $\Phi_{N_{\sigma\sigma'}}=H_{\sigma\sigma'}^{cv}(\vec{k}=0)$ as a,
	possibly spin-dependent, nematic order parameter.
	Similarly since $s_0\tau_0$,$s_0\tau_x$,$s_0\tau_z$,$s_x\tau_y$,$s_y\tau_y$, and $s_z\tau_y$ are 
	time-reversal invariant, it follows that when time-reversal symmetry is intact 
	the quasiparticle Hamiltonian at wave vector $\vec{k}=0$ must satisfy
	$H_{\uparrow\uparrow}^{ss}=H_{\downarrow\downarrow}^{ss}$, 
	$H_{\uparrow\uparrow}^{ss'} = [H_{\downarrow\downarrow}^{ss'}]^*$,
	$H_{\sigma\sigma'}^{ss'}= - H_{\sigma\sigma'}^{s's}$,
	and $H_{\sigma\sigma'}^{ss}=0$, where $s \neq s'$ and $\sigma \neq \sigma'$.
	We can define four corresponding order parameters that characterize different ways 
	in which the system can break time-reversal symmetry:
	$\Phi_{1}=H_{\uparrow\downarrow}^{cv}(\vec{k}=0)+H_{\uparrow\downarrow}^{vc}(\vec{k}=0),
	\Phi_{2}=H_{\uparrow\uparrow}^{cc}(\vec{k}=0)-H_{\downarrow\downarrow}^{cc}(\vec{k}=0),
	\Phi_3=Re(H_{\uparrow\uparrow}^{cv})(\vec{k}=0)-Re(H_{\downarrow\downarrow}^{cv})(\vec{k}=0)+
	Im(H_{\uparrow\uparrow}^{cv}(\vec{k}=0))+Im(H_{\downarrow\downarrow}^{cv}(\vec{k}=0))$,
	and $\Phi_4=H_{\uparrow\downarrow}^{cc}(\vec{k}=0)$.
	
	{\em Phase diagram.}---
	\begin{figure}[t]
		\includegraphics[width=1\columnwidth]{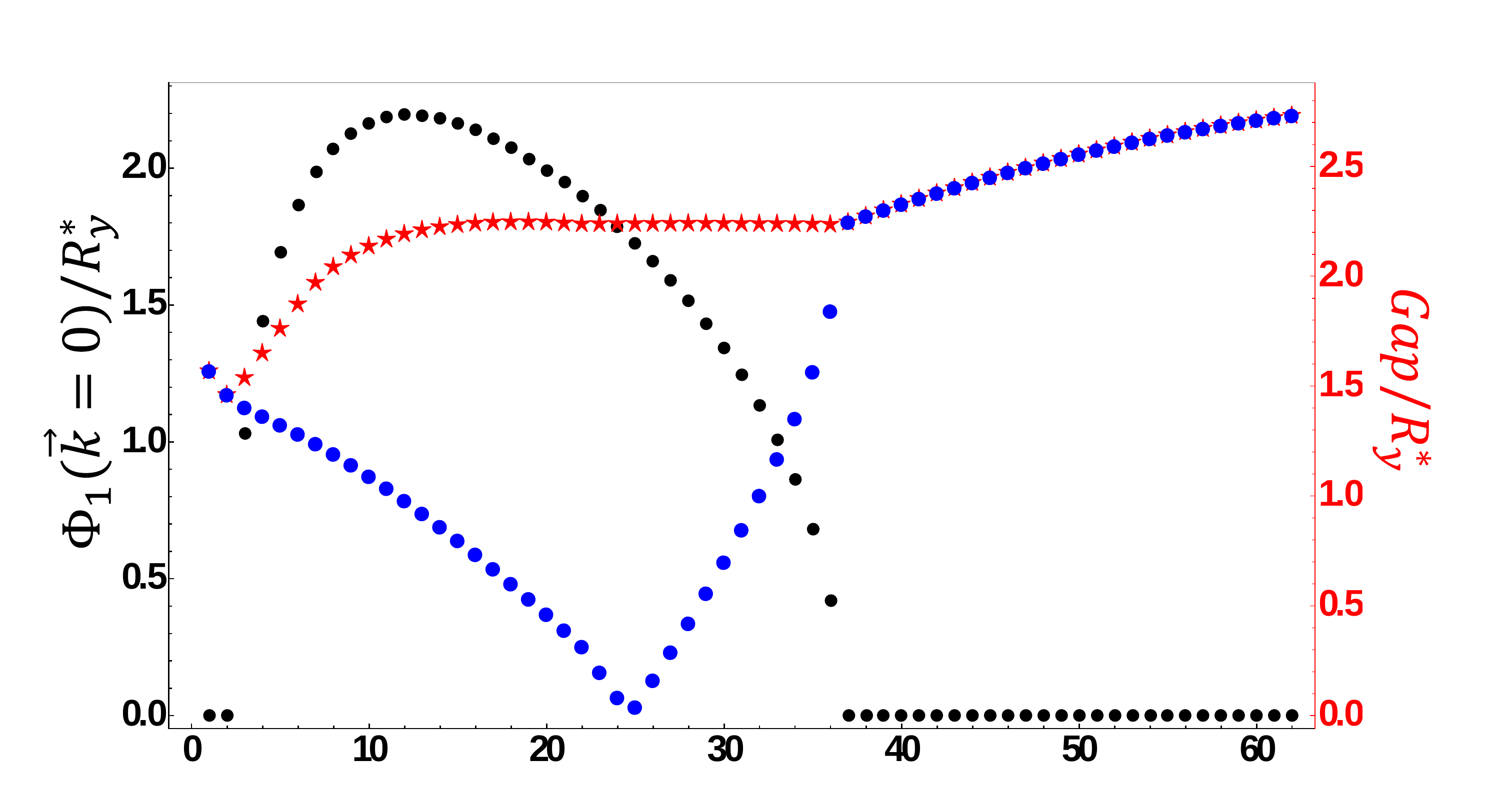}
		\caption{(Color online) 
			Time-reversal symmetry-breaking order parameters $\Phi_1 $ (black dots, left axis) 
			and quasiparticle gaps (red stars, right axis)
			along the gray dashed line in Fig.~\ref{Fig:PhaseDiagram}, 
			which passes through the TRSB nematic insulator state.
			For comparison, the blue squares show the quasiparticle gaps in the time-reversal symmetry-preserving  
			nematic insulator state discussed in the main text, which has higher energy.
		}
		\label{Fig:circle}
	\end{figure}
	The phase diagram in Fig.~\ref{Fig:PhaseDiagram} was constructed
	by identifying the lowest energy solution of Eq. ~\ref{eq:MF} over a range of 
	$A$ and $E_{g}$ values\cite{Du2011, Du2015, Du2017},
	fixing other model parameters at values appropriate for 
	InAs/GaSb QWs: electron-hole layer separation $d=0.3 a_B^*\sim100{\AA}$,
	$m_e=0.023m_0$, $m_h=0.4m_0$, and $\epsilon\sim15\epsilon_0$\cite{GaSb,InAs}.
	($a_B^*\sim365{\AA}$ and $\text{Ry}^*\sim1.3meV$.) 
	In InAs/GaSb systems, the energy gap $E_g$ can be varied by 
	changing quantum well widths, and tuned {\it in situ} with external gates\cite{Du2011,Du2015}.
	The band hybridization parameter $A$ can be varied by inserting AlSb barrier layers between
	the InAs electron layer and the GaSb hole layer\cite{Du2017}.
	
	The black diamond on the vertical axis ($A=0$ line)
	in Fig.~\ref{Fig:PhaseDiagram} marks the 
	point at which the band gap is reduced to the spatially indirect 
	exciton binding energy.  
	The $s$-wave exciton condensate state\cite{Zhu1995,Wu&Xue2015}
	which forms at this point when $A=0$ establishes spontaneous coherence between 
	bands that is peaked at $\vec{k}=0$, is independent of momentum orientation $\theta_{k}$,
	and has an energy that is invariant under independent spin reorientations in either layer.  
	We find that at finite $A$ the ground state prefers that  
	coherence be established between opposite spins, and that the two independent 
	opposite-spin coherence parameters $H_{\uparrow\downarrow}^{cv}(\vec{k}=0)$
	and $H_{\uparrow\downarrow}^{vc}(\vec{k}=0)$ prefer to have the same sign, 
	breaking rotational and time-reversal symmetry.
	This arrangement minimizes the frustration between $s$-wave exciton 
	condensation and single-particle interband coupling
	that is proportional to $\exp(i\theta_{\vec{k}})$ and diagonal in spin.
	The occupied quasiparticles have band-spin spinors of the form 
	$(u_{\vec{k}},v_{\vec{k}}e^{-\mathrm{i}\theta_{\vec{k}}},
	u_{\vec{k}}e^{-\mathrm{i}\theta_{\vec{k}}},v_{\vec{k}})^T$ allowing their projection onto a 
	definite spin to have p-wave interband coherence, while retaining opposite spin coherence
	that is independent of $\theta_{k}$.
	Interband coherence at $\vec{k}=0$ breaks the BHZ model's
	rotational symmetries.      
	In the simplified spinless two-band model 
	(see supplemental material), the frustration between $s$-wave excitons and 
	$p$-wave contributions to the band Hamiltonian is resolved in momentum space 
	by moving the vortex in $H^{cv}$ away from $\vec{k}=0$.
	Adding the spin degree of freedom enables 
	a resolution of the frustration between interaction and band terms in the Hamiltonian
	that is simpler and more elegant  than in the spinless case discussed in the Supplementary Material.  
	
	We do find solutions of the mean-field equations with 
	$H_{\uparrow\downarrow}^{cv}(\vec{k}=0)= - H_{\uparrow\downarrow}^{vc}(\vec{k}=0)$,
	preserving time-reversal symmetry, but these always have higher total energy than the  
	time-reversal symmetry-breaking(TRSB) $\Phi_1 \ne 0$ solutions.
	TRSB states are energetically preferred because they provide a continuous phase transition path between
	ordinary insulator states and QSHI states along which the gap is not required to vanish \cite{Liu2008,Hyart2014,Sau2016}.
	The quasiparticle Hamiltonian of the time-reversal symmetry-preserving nematic state
	has the form $\xi_{\vec{k}}s_0 \tau_z+A k_xs_z\tau_x-A k_ys_0\tau_y+X s_y \tau_y $, 
	where $X$ is an exchange energy, and therefore a 
	gap $2\sqrt{\xi_{\vec{k}}^2+(Ak_y-X)^2}$ that vanishes when $\xi_{\vec{k}}=0$ and 
	$A k_y = X$.  These conditions are satisfied along a 
	line in phase space that cannot be avoided in transiting between normal and QSHI states,
	as illustrated by the gap closing phase transition(blue squares) in Fig.~\ref{Fig:circle}.
	On the other hand, the TRSB state has a mean-field Hamiltonian of the form 
	$\xi_{\vec{k}}s_0 \tau_z+A k_xs_z\tau_x-A k_ys_0\tau_y+X s_x \tau_x $,
	implying a gap, $2\sqrt{\xi_{\vec{k}}^2+(Ak)^2+X^2}$, that needs not to vanish.

	
	Fig.~\ref{Fig:PhaseDiagram} identifies five distinct phases with different
	order parameters and band topologies.  In addition to the normal and QSHI phases 
	of the bare bands, three interaction-induced phases appear
	all of which break time-reversal symmetry.
	The TRSB nematic insulator has nonzero values for $\Phi_{N}$ and $\Phi_1$; 
	the $XY$ insulator has spontaneous transverse spin polarization in addition so that 
	$\Phi_{N}$, $\Phi_1$, and $\Phi_4$ are all non-zero; the QAHI state has a 
	nonzero value of order parameter $\Phi_2$, but is not nematic.  Its presence close to the line
	along which the interaction renormalized band gap 
	vanishes is closely related to the heavily studied instabilities of massless\cite{MacDonald2012,Semenoff2012}
	two-dimensional Dirac models at strong interactions.  
	[$A/(Ry^{*} a_B^{*})  = 2 [e^2/(\hbar v\epsilon)]^{-1}$ where $v=A/\hbar$ is the 
	band velocity at $\xi_{\vec{k}=0}=0$.]
	The normal insulator and the QSHI preserve time-reversal and rotational symmetry,
	and differ only in the sign of the renormalized band gap at $\vec{k}=0$.
	
	
	At large values of $A$ the NI/QSHI transition is not altered by interactions.  
	At intermediate values of $A$, we find that the NI/QSHI transition proceeds via an 
	intermediate QAHI state that is separated from both NI and QSHI states by 
	first-order phase transitions, similar to the behavior predicted
	by dynamic mean-field theory for Hubbard model systems\cite{Sangiovanni2015} and 
	by mean-field theory for interacting Kane-Mele Hubbard models\cite{Wang2018}.  
	The QAHI phase is characterized by a $U s_z \tau_z$ mean-field term and has a $\Phi_2$ TRSB order parameter. 
	To characterize the topological properties of the various different phases  
	we perform a continuum model version of a Wilson loop\cite{Yu2011,Fang2015} calculation for 
	the two occupied bands. We evaluate the non-Abelian
	$2\times2$ Berry connection matrix $F_{i,i+1}^{m,n}=\braket{u_i^m |u_{i+1}^n}$
	along square loops of different perimeters surrounding the momentum space origin.
	Then we construct a matrix $D$ by finding the product of all $F$s along the square path labeled by $k$, equal
	to half of the square's edge.  These matrices have 
	two eigenvalues and phase angles $\theta_{k}$.
	The change in the sum of the $\theta_{k}$ values between $k=0$ and a finite value of $k$ 
	is\cite{Yu2011,Fang2015} equal to the integral of the momentum space Berry curvature 
	over the enclosed area.  Because band inversion occurs only near $k=0$, 
	we can identify the topological properties of quasiparticle bands from these
	small $k$ continuum model calculations.  
	
\begin{figure}[t]
		\includegraphics[width=1\columnwidth]{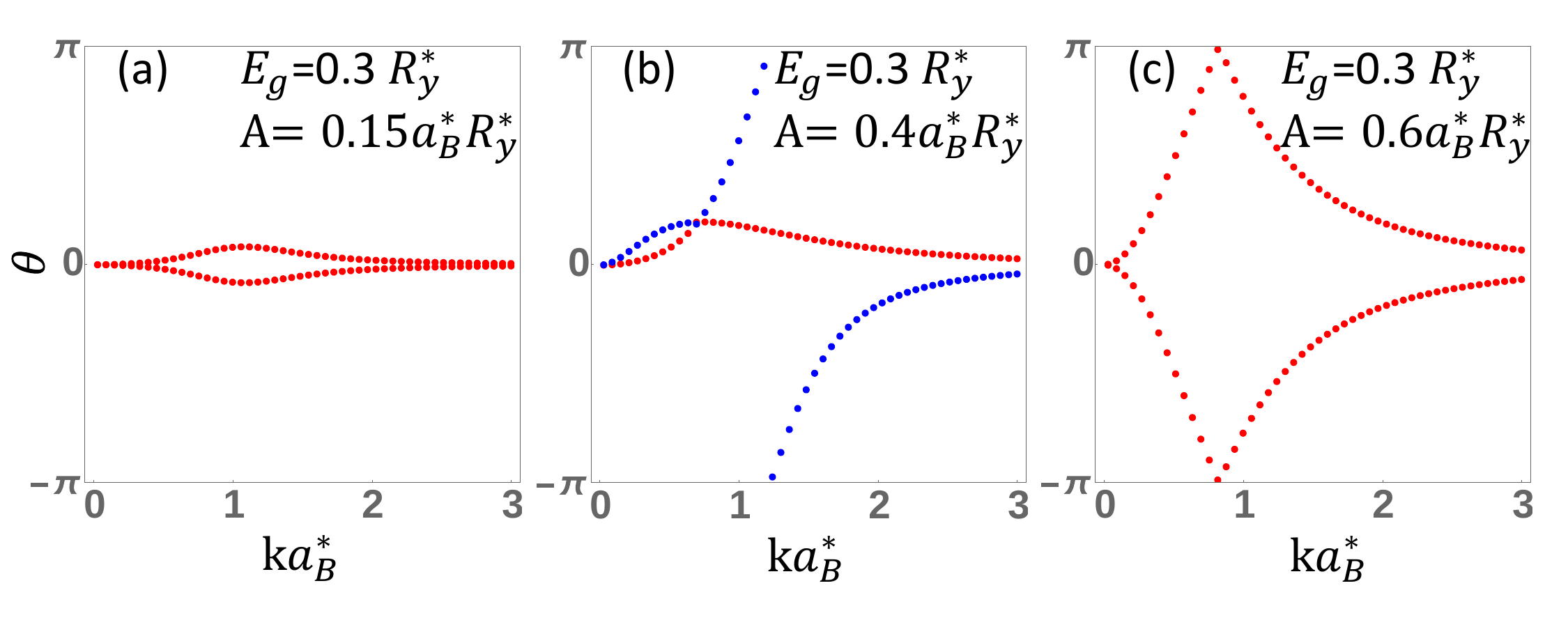}
		\caption{(Color online) 
			Phase angle $\theta$ profiles at the three different phase diagram points marked in 
			Fig.~\ref{Fig:PhaseDiagram}.  For the left (TRSB nematic insulator) and right (QSHI) states the quasiparticle bands 
			are doubly degenerate.  In the middle panel (QAHI state) the blue and red dots distinguish the phase angles
			of the two occupied bands. (a) TRSB nematic insulator state with topologically trivial bands, (b) QAHI with one
			nonzero Chern number band, and (c) QSHI with two opposite nonzero Chern number bands.
		}
		\label{Fig:Wilsonloop}
	\end{figure}

	In Fig.~\ref{Fig:Wilsonloop} we plot typical $\theta$ profiles for
	TRSB nematic insulator, QAHI, and QSHI phases. Fig.~\ref{Fig:Wilsonloop}(a) shows that 
	the TRSB nematic insulator is topologically trivial, with two winding number zero bands.
	The $\theta_{k}$ profiles of the $XY$ insulator, and normal
	insulator states (not shown) are similar to those of the TRSB nematic insulator state.  
	Fig.~\ref{Fig:Wilsonloop}(b) shows that 
	the QAHI state is topologically nontrivial with one band winding the cylinder once,
	corresponding to total Chern number equal to 1.
	Similarly Fig.~\ref{Fig:Wilsonloop}(c) demonstrates 
	the topological nontrivial $Z_2=1$ behavior expected for a QSHI,
	with two bands winding the cylinder once in opposite directions.
	The topology can be identified from these Chern number calculations 
	because up and down spin sectors are decoupled.  
	
	%
	
	{\em Discussion.}---
	The BHZ model applies to HgTe/CdTe and InAs/GaSb quantum well systems.
	In the former case the electron and hole bands are strongly coupled because they 
	both reside in HgTe.  The dimensionless band-coupling 
	parameter $A/(Ry^{*}a_{B}^*)$ is therefore large\cite{Liu2013} and interactions are unimportant.
	The phase diagram in Fig.~\ref{Fig:PhaseDiagram} can
	be fully explored experimentally in InAs/GaSb systems 
	by inserting AlSb layers between the InAs and GaSb to vary $A$.  
	Indeed important progress has already been achieved in recent studies of the no AlSb\cite{Du2015} 
	and thick AlSb\cite{Du2017} ($A =0$) limits.
	In the absence of AlSb, it was shown\cite{Du2015} that the band gaps in the 
	QSHI state are larger than expected on the basis of single-particle physics 
	alone, as predicted by our mean-field calculations, 
	and that they survive in-plane magnetic fields that are expected to 
	suppress single-particle contributions.
	Qualitatively, in-plane 
	magnetic fields have an effect similar to reducing $A$.  
	The observation that the gap does not vanish even as its single-particle 
	support is removed is consistent with our findings.  Further work will be necessary to 
	determine whether or not the resulting state has the broken time-reversal and 
	rotational symmetry that we expect in the small $A$ limit.
	
	
	It is important to recognize that mean-field theory can err both quantitatively 
	and qualitatively. For example, the stability region of the QAHI state along the 
	$\xi_{\vec{k}=0}=0$ line is expected\cite{MacDonald2012,Semenoff2012} to be shifted toward stronger interactions (smaller $A$) 
	by quantum fluctuations, and could potentially be preempted by the TRSB nematic insulator state.
	
	The presence of a QAHI state can be established experimentally by 
	performing nonlocal transport measurements, similar to those\cite{Konig2007,Du2011} used to 
	establish the QSHI state to establish that edge states have chiral rather than helical character.  
	The appearance of an exciton condensate state along  
	the $A=0$ line, where the physics is simplified by binding of electrons and holes into 
	bosonic excitons, is certain, but the density at which the Mott transition
	occurs\cite{Liu1998,DePalo2002,Nikolaev2008,Asano2014} is difficult to estimate accurately. 
	Since it can be viewed as an exciton condensate that 
	is weakly perturbed by band hybridization, the appearance of a
	TRSB nematic insulator state is also certain, but its persistence in the upper left-hand side of our phase diagram
	(Fig.~\ref{Fig:PhaseDiagram}) where it competes with paramagnetic metallic states is uncertain.
	The presence of a TRSB nematic insulator state can be established by performing counterflow 
	experiments\cite{Su2008} and by demonstrating the absence of edge states.
	(A similar nematic phase has been proposed in the vicinity of 
	quantum anomalous Hall states\cite{Sun2009, Cook2014}.)
	Very recent experimental studies\cite{Du2017} have demonstrated that a gapped state, presumably
	the exciton condensate, is still present at $n_{ex} a_{B}^{*2} \approx 0.03$.
	These findings suggest that the full region of the phase diagram 
	in Fig.~\ref{Fig:PhaseDiagram} is open to experimental study.
	Recently, a new type of QSHI has been discovered experimentally\cite{Cobden2017,Shen2017,Wu2018},
	which is described by band models\cite{Qian2014} that are distinguished from the BHZ model 
	studied here mainly by large anisotropies and also subject to interaction-induced broken symmetries.
	%
	%
	
	This work was primarily supported by the U.S. Department of Energy, Office of Science,
	Basic Energy Sciences under Award No. DE-FG02-ER45958 and by the Welch Foundation under Grant No. TBF1473.

\appendix
\section{Supplemental Material}
	
	In this supplemental material, we will discuss a simple one-spin two-band case on interaction-induced
	nematic insulators near quantum spin Hall phase transitions.
	The single particle Hamiltonian for down spin in BHZ model is:
	\begin{equation}
	\label{eq:single_Hamiltonian_2}
	H_{0,\downarrow}=\begin{pmatrix}
	\frac{\hbar^2 k^2}{2m_e}+E_c & -Ak_- \\
	-Ak_+ & -\frac{\hbar^2 k^2}{2m_h}+E_v \\
	\end{pmatrix},
	\end{equation}
	where $k_{\pm}=k_x\pm \mathrm{i} k_y$, $A$ is the band couping strength, $m_{e(h)}$ is the electron(hole)
	effective mass.
	
	The Coulombic
	electron-electron interaction Hamiltonian
	\begin{equation}
	\label{eq:interaction_Hamiltonian_2}
	\hat{H}_I=\frac{1}{2S}\sum_{ss'}\sum_{\vec{k},\vec{k}',\vec{q}}V^{ss'}(\vec{q})
	a_{s \vec{k}}^{\dagger}a_{s' \vec{k}'}^{\dagger}
	a_{s' \vec{k}'+\vec{q}}a_{s \vec{k}-\vec{q}} \, 
	\end{equation}
	where $S$ is the two dimensional system area, conduction or valence bands are 
	labeled by $s(s')=c$ or $v$, $a_{s \vec{k}}^{\dagger}$ and $a_{s \vec{k}}$ are 
	quantum well (QW) conduction(valence) band electron creation and annihilation operators, 
	$V^{cc}(\vec{q})=V^{vv}(\vec{q})=V(\vec{q})=2\pi e^2/(\epsilon q)$, and
	$V^{cv}(\vec{q})=V^{vc}(\vec{q})=U(\vec{q})=V(\vec{q}) e^{-qd}$.  We are interested in the 
	properties of this interacting electron system as the band gap $E_{g}=E_c-E_v$ 
	closes and changes sign.
	
	When $A$ vanishes, the model reduces to that of the two-dimensional excitonic insulator
	problem,\cite{Zhu1995,Lozovik1996,Fernandez1996,Vina1999,Combescot2008a,Combescot2008b,Butov2012,Perali2013,Wu&Xue2015}
	in which electron-electron interactions play a central role.
	For gaps larger than the exciton binding energy, the ground state at $A=0$ is a trivial band insulator.
	For gaps slightly smaller than the exciton binding energy, there is a continuous phase 
	transition to a ground state that is still insulating but populated by a condensate of excitons with weakly 
	repulsive interactions.  
	At still smaller gaps, there is a first order phase transition\cite{Liu1998,DePalo2002,Nikolaev2008,Asano2014} 
	to a conducting state with free 
	electrons and holes.  The characteristic length scale of
	the excitonic insulator problem is the effective Bohr radius $a_B^*=\epsilon \hbar^2  /(me^2)$, and the 
	characteristic energy scale is the effective Rydberg $\text{Ry}^*=e^2/(2\epsilon a_B^*)$.
	(Here $m=m_em_h/(m_e+m_h)$ is the excitonic reduced mass.)  When $A$ is very much larger than 
	$Ry^* a_B^*$, interactions play an unimportant role.  We explain below how the excitonic insulator physics 
	evolves with increasing $A$ into a renormalized version of single-particle 
	normal insulator/Quantum spin Hall insulator (NI/QSHI) phase-transition physics, and why the crossover is punctuated by a 
	nematic state with broken rotational symmetry.

	{\em Microscopic Mean-Field Theory:}---
	Because spin-orbit coupling terms that mix spins are absent in the 
	BHZ model, we focus initially on a single block.  
	The Hartree-Fock mean-field Hamiltonian for down spins,
	\begin{equation}
	\label{eq:MF_2}
	\hat{H}_{MF}=\sum_{\vec{k}} (a_{c\vec{k}}^{\dagger}, a_{v\vec{k}}^\dagger)
	\begin{pmatrix}
	\zeta_{\vec{k}}+\xi_{\vec{k}} & -\Delta_{\vec{k}} \\
	-\Delta_{\vec{k}}^* & \zeta_{\vec{k}}-\xi_{\vec{k}} \\
	\end{pmatrix}
	\begin{pmatrix} a_{c\vec{k}}\\a_{v\vec{k}}\end{pmatrix}\\.
	\end{equation}
	Here $\zeta_{\vec{k}}=\hbar^2k^2[1/(4m_e)-1/(4m_h)]$ accounts for the mass difference 
	between conduction and valence bands which plays a very minor
	role in selecting between insulating many-particle ground states and is dropped below.  
	The difference ($\xi_{\vec{k}}$) between conduction and valence band energies 
	and the band coupling amplitude ($\Delta_{\vec{k}}$) are both renormalized by interactions and
	determined by solving the following self-consistent field equations:
	\begin{equation}
	\label{eq:SC_2}
	\begin{split}
	\xi_{\vec{k}} &= \frac{\hbar^2k^2}{4m} + \frac{E_{gap} + 4\pi e^2 n_{c} d/\epsilon}{2} \\
	& -\frac{1}{2S}\sum_{\vec{k}'}V(\vec{k}-\vec{k}')(1-\xi_{\vec{k}'}/E_{\vec{k}'}),\\
	&\Delta_{\vec{k}}=\frac{1}{2S}\sum_{\vec{k}'}U(\vec{k}-\vec{k}')\frac{\Delta_{\vec{k}'}}{E_{\vec{k}'}}+
	Ak_-,
	\end{split}
	\end{equation}
	where $E_{\vec{k}}=\sqrt{\xi_{\vec{k}}^2+\abs{\Delta_{\vec{k}}}^2}$,
	and $d$ is the vertical separation between electron and hole layers.  
	Note that the important model Hamiltonian 
	parameter $E_{gap}=E_c-E_v$ is equal to the quasiparticle energy gap at $\bm{k}=0$ only
	when $\Delta_{\vec{k}} = 0$ so that the conduction band is completely 
	empty and the valence band full.  In Eq.~\ref{eq:SC_2}
	\begin{equation}
	n_{c}=\frac{1}{2S}\sum_{\vec{k}}(1-\xi_{\vec{k}}/E_{\vec{k}})
	\end{equation}
	is the charge density in the conduction band layer.  
	
	$\Delta_{\vec{k}}=\abs{\Delta_{\vec{k}}} \exp(\mathrm{i}\phi_{\vec{k}})$
	in Eq.~\ref{eq:MF_2} is a complex function of $\vec{k}$.
	In the $A=0$ excitonic insulator state $\Delta_{\vec{k}}$ is independent of
	$\theta_{\vec{k}}=\arctan(k_y/k_x)$.  In the large $A$ limit, however, 
	$\Delta_{\vec{k}}$ has the same $\theta_{\vec{k}}$-dependence as 
	$Ak_- = A k \exp(-\mathrm{i}\theta_{\vec{k}})$.
	The $k$ and $\theta_{\vec{k}}$ dependence at intermediate 
	values of $A$ minimizes the total energy per area,
	\begin{widetext}
		\begin{equation}
		\label{eq:totalenergy_2}
		\epsilon=\frac{1}{2S}\sum_{\vec{k}}\bigg[(\frac{\hbar^2k^2}{4m}+ \frac{E_{gap}}{2} +\xi_{\vec{k}})
		(1-\frac{\xi_{\vec{k}}}{E_{\vec{k}}})
		-\frac{\abs{\Delta_{\vec{k}}}^2+Ak\abs{\Delta_{\vec{k}}} \cos(\theta_{\vec{k}}+\phi_{\vec{k}})}{E_{\vec{k}}}\bigg].
		\end{equation}
	\end{widetext}
	
	\begin{figure}[b]
		\includegraphics[width=1\columnwidth]{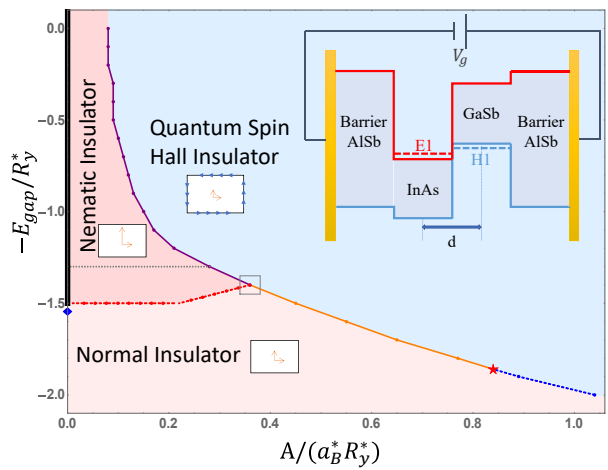}
		\caption{(Color online) 
			Mean-field phase diagram of a model quantum spin Hall insulator (QSHI) as a function of band inversion
			parameter $E_{gap}$ and band coupling strength parameter $A$.  
			The blue square on the $A=0$ axis separates a normal insulator state and 
			an exciton condensate (XC) state which is present where the axis is bold.  
			The light blue area denotes QSHI phase
			where edge state for one spin exists and bulk transport is isotropic. The dark pink
			area denotes nematic insulator phase where no edge state exists and bulk transport is
			anisotropic due to rotational symmetry breaking. The light pink area denotes normal
			insulator phase where no edge state exists and bulk transport is isotropic.
			The purple line describes first order phase transition boundary between 
			nematic insulator and QSHI. The red dashed line describes second order phase transition
			boundary between normal insulator and nematic insulator. The orange line describes 
			first order NI/QSHI phase transition boundary. The blue dashed
			line describes second order NI/QSHI phase transition boundary. 
			Red star denotes critical value where NI/QSHI phase transition becomes
			continuous. The gray square denotes the parameter space where we calculate off-diagonal
			and diagonal terms at $\vec{k}=0$ in Fig.~\ref{Fig:circle_2}.
			The inset shows a schematic band diagram for 
			the AlSb/InAs/GaSb/AlSb QW system to which the model corresponds most closely.  
			In this system electron and hole layers are spatially separated and the band inversion parameter $E_{gap}$ 
			can be tuned by adjusting gate voltage that apply electric fields across AlSb barriers.
		}
		\label{Fig:PhaseDiagram_2}
	\end{figure}
	
	{\em Phase Diagram}---The phase diagram in Fig.~\ref{Fig:PhaseDiagram_2} was constructed
	by solving Eqs~\ref{eq:SC_2} over a range of $A$ and $E_{gap}$ values,\cite{Du2011} 
	fixing other model parameters at values appropriate for 
	InAs/GaSb QWs: electron-hole layer separation $d=0.3 a_B^*\sim100{\AA}$,
	$m_e=0.023m_0$, $m_h=0.4m_0$, and $\epsilon\sim15\epsilon_0$\cite{GaSb,InAs}.
	These values set $a_B^*\sim365{\AA}$ and $\text{Ry}^*\sim1.3meV$. 
	In InAs/GaSb systems, the energy gap $E_g$ can be varied by changing quantum well widths, and tuned {\it in situ} with external gates.  The band hybridization parameter $A$ can be varied by inserting AlSb layers between
	the InAs electron layer and the GaSb hole layer.

	The blue square on the vertical axis $A=0$ in Fig.~\ref{Fig:PhaseDiagram_2} marks the 
	point at which the band gap is reduced to the spatially indirect 
	exciton binding energy.  Our main finding is that the exciton condensate state\cite{Zhu1995,Wu&Xue2015}
	which forms at this point when $A=0$ and induces inter-band coherence that is strongest 
	at $\vec{k}=0$ and phase $\phi_{\vec{k}}$ that is independent of momentum orientation $\theta_{\vec{k}}$, is
	only weakly perturbed by band hybridization; $\phi_{\vec{k}}$ is independent of $\theta_{\vec{k}}$ in the 
	$A=0$ ground state because only $s$-wave excitons are energetically allowed at relevant $E_{g}$ values.  
	We find that the interaction contribution to $\Delta_{\vec{k}}$ initially changes gradually
	with $A$.  When the single-particle contribution is added $|\Delta_{\vec{k}}|$ and the quasiparticle energy 
	$E_{\vec{k}}$ are no-longer independent of momentum orientation $\theta_{\vec{k}}$, inducing anisotropy 
	in all electronic properties.  Indeed, because the $\vec{k}=0$ electron and hole states have different 
	angular momentum, hybridization between them that does have a constant value of
	$\theta_{\vec{k}}+\phi_{\vec{k}}$ must break the BHZ model's rotational symmetries.
	We identify the small $A$ state at gaps that are smaller than the 
	exciton binding energy as a nematic insulator.

	\begin{figure}[t]
		\includegraphics[width=1\columnwidth]{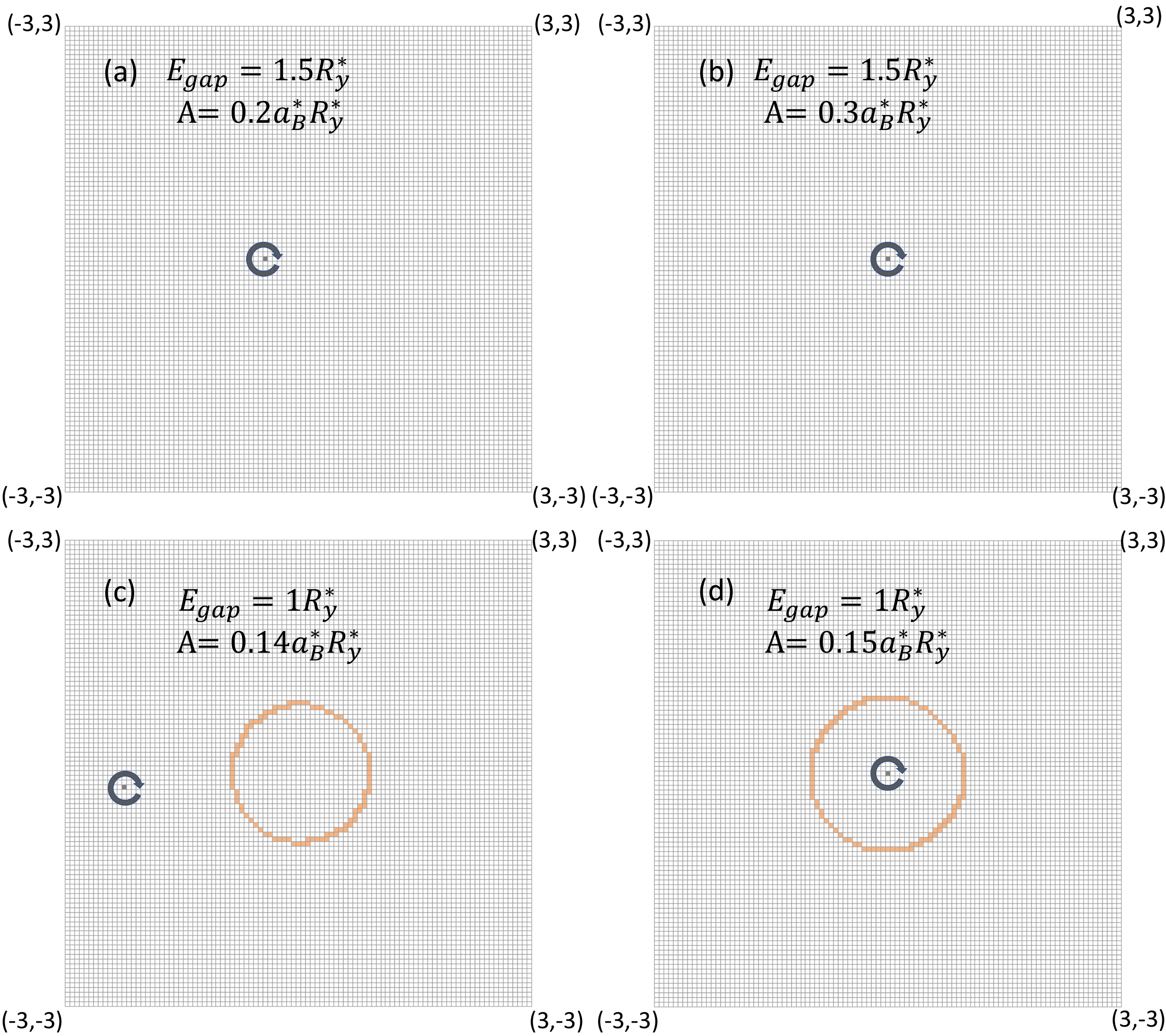}
		\caption{(Color online) 
			Plots of vortex points and zero-lines on the momentum grid under different circumstances
			where black plaquettes denote vortex points surrounded by a schematic arrow 
			and orange plaquettes denote $\xi_{\vec{k}}=0$. 
			(a) Nematic insulator state at ($\mu=-1.5,A=0.2$); (b) Normal insulator state at ($\mu=-1.5,A=0.3$);
			(c)Nematic insulator state at ($\mu=-1,A=0.14$); (d) QSHI state at ($\mu=-1,A=0.15$).
		}
		\label{Fig:vortex}
	\end{figure}
	
	\begin{figure}[t]
		\includegraphics[width=1\columnwidth]{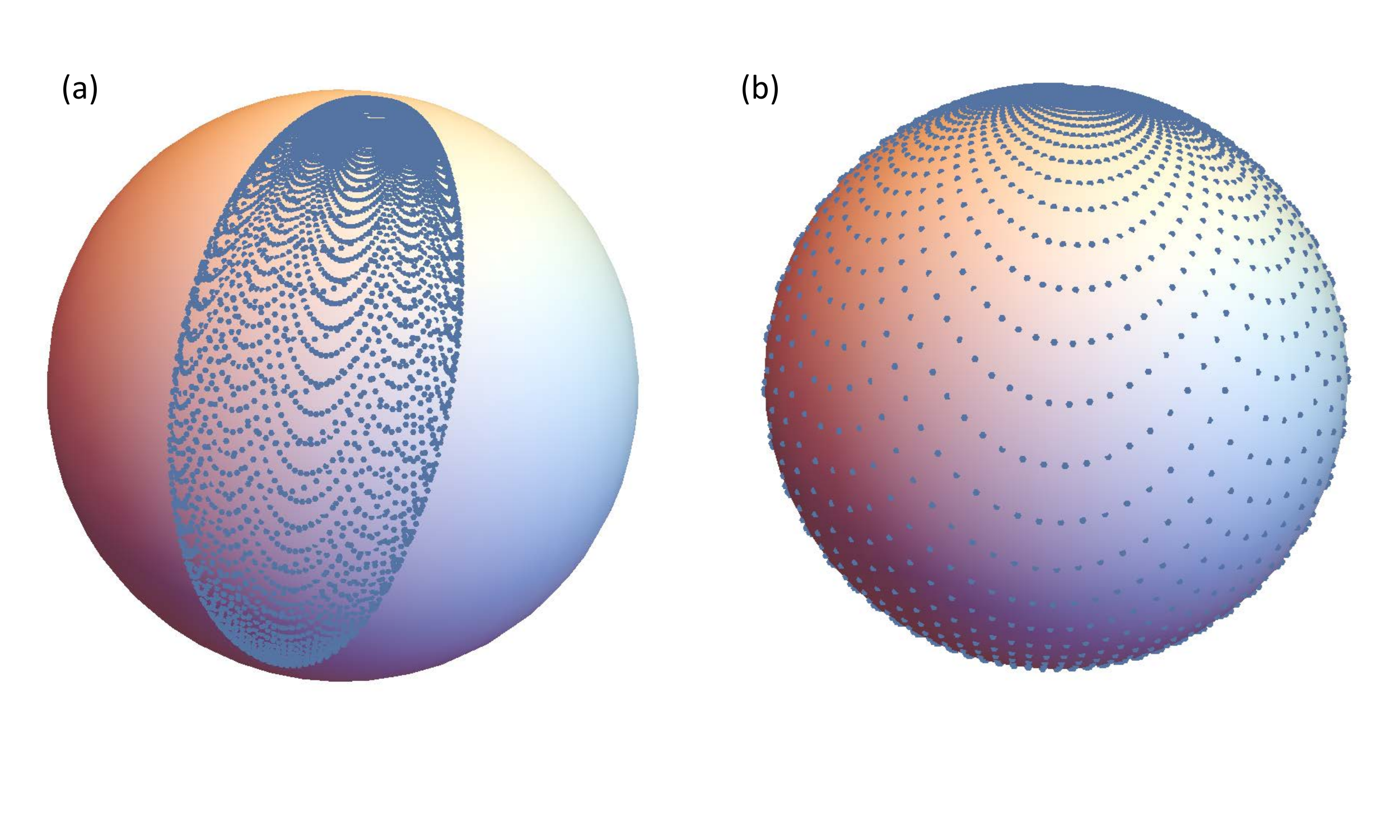}
		\caption{(Color online) 
			Plots of Bloch spheres. 
			(a) Nematic insulator state at ($\mu=-1,A=0.14$); (b) QSHI state at ($\mu=-1,A=0.15$).
		}
		\label{Fig:blochsphere}
	\end{figure}
	
	The relationship of the nematic insulator to the NI/QSHI phase transition is best addressed by 
	considering the Chern index of the spin-projected quasiparticle bands. 
	Because $\Delta_{\vec{k}}$ is complex, it will vanish at isolated points in momentum space.
	Because the single-particle contribution dominates at large $k$, the line integral of its phase
	derivative around a large diameter circle must equal the single-particle value $ - 2 \pi$.  We conclude that 
	$\Delta_{\vec{k}}$ always has isolated vortex at some value of $\vec{k}$.  Whenever 
	the vortex is not at $\vec{k}=0$, rotational symmetry is broken.  
	Similarly the renormalized band gap function $\xi_{\vec{k}}$ is real and therefore 
	can vanish along a line, referred to below as the zero-line, in momentum space.  
	By taking note of the different sense of 
	dispersion in the conduction and valence bands, we see that the occupied quasiparticle band 
	has a non-zero Chern number when the zero-line is present and, the vortex is enclosed 
	by the zero-line.  To identify vortex positions, we calculated the phase winding around each momentum
	mesh plaquette on our grid.  Plaquettes with $2\pi$ phase winding are shaded black in Fig.~\ref{Fig:vortex}
	and surrounded by an arrow indicating the sense of vorticity. 
	The nematic state in Fig.~\ref{Fig:PhaseDiagram_2} is distinguished by
	$\Delta_{\vec{k}}$ vortices that are located away from $\vec{k}=0$ and outside the zero-line
	when one is present, as illustrated in Fig.~\ref{Fig:vortex}(a) and (c). 
	The normal insulator and QSHI states both have $\Delta_{\vec{k}}$ vortices at $\vec{k}=0$,
	as illustrated in Fig.~\ref{Fig:vortex}(b) and (d)), 
	but a zero line is present only for the QSHI state in Fig.~\ref{Fig:vortex}(d).  
	
	States with vortices
	away from $\vec{k}=0$, but inside the zero-line, which would be nematic
	topological insulator states, did not appear in our mean-field calculations.
	Instead, as $A$ increases from a point inside the nematic insulator region
	the ground state vortex position always jumps discontinuously
	from outside the zero line to the origin. 
	This vortex position jump demonstrates that the nematic insulator to QSHI 
	phase transition is first order.  
	The critical value of $A$ at which the transition occurs 
	decreases when $E_{gap}$ decreases 
	because the zero-line becomes moves to larger 
	momentum magnitude as the bare gap decreases.

	In Fig.~\ref{Fig:blochsphere} we illustrate the relationship the connection between the
	vortex positions relative to zero-line and band Chern numbers by providing a  
	Bloch sphere representation of the dependence of occupied quasiparticle state 
	on momentum.  We have mapped states on our momentum-space grid are mapped to the Bloch unit sphere
	$(\sin(\theta_{B}) \cos(\phi_{B}),\sin(\theta_{B})\sin(\phi_{B}),\cos(\theta_{B}))$
	using $\cos(\theta_{B}) = \xi_{\vec{k}}/E_{\vec{k}}$ and $\cos\phi_{B} = -\frac{\abs{\Delta_{\vec{k}}}\cos(\phi_{\vec{k}})}{E_{\vec{k}}}$.
	In a topological nontrivial state, the mapping from momentum space 
	covers the whole Block sphere, as illustrated in Fig.~\ref{Fig:blochsphere}(b). 
	A zero-line is necessary for points to cross the equator and only a vortex inside a zero-line 
	can cover the entire equator.
	
	\begin{figure}[t]
		\includegraphics[width=1\columnwidth]{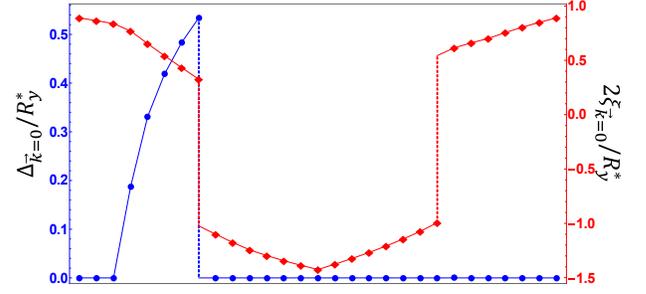}
		\caption{(Color online) 
			Plots of $\Delta_{\vec{k}=0}$ and $2\xi_{\vec{k}=0}$ at different parameter space.
			Blue dot denotes off-diagonal terms at origin, and only nematic insulator phase 
			has non-zero value. Red square denotes diagonal difference terms at origin, and
			phase transitions of both nematic insulator/QSHI and NI/QSHI are
			first order. Phase transition between nematic to normal insulator is second order.
		}
		\label{Fig:circle_2}
	\end{figure}
	
	We take a careful look at phase transition areas (a gray square) where three phase transition lines touch together as 
	shown in Fig.~\ref{Fig:PhaseDiagram_2}. To determine whether the phase transitions between different phases
	are continuous or not, we plot order parameter $\Delta_{\vec{k}=0}$ and renormalized gap $2\xi_{\vec{k}=0}$
	at various parameters along the gray square. In Fig.~\ref{Fig:circle_2}, blue dots represent nematic order parameters
	$\Delta_{\vec{k}=0}$ and only nematic insulator phase has non-zero value. Red squares represent differences in diagonal
	terms at $\vec{k}=0$ and QSHI requires the renormalized gap to be negative. Along the gray square, first phase transition
	normal insulator/nematic insulator is second order because both $\Delta_{\vec{k}=0}$ and $2\xi_{\vec{k}=0}$
	changes continuously with discontinuous slope. Then second phase transition nematic insulator/QSHI is first order because
	of a discontinuous jump in both terms guided by dashed vertical lines. Third phase transition QSHI/normal insulator is
	also first order because of a discontinuous jump in renormalized gap. This is unexpected for a non-interacting topological
	phase transition where the gap between conduction and valence band closes continuously. But due to the long-range
	Coulomb interaction, even NI/QSHI phase transition at small band gap becomes first order. When band gap is very large,
	we expect the NI/QSHI transition to be continuous again as Coulomb interaction becomes irrelevant. This critical band
	gap value has been marked as red star in Fig.~\ref{Fig:PhaseDiagram_2}.

\bibliography{QWEXC.bib}{}
\bibliographystyle{apsrev4-1}

\end{document}